\documentclass[aps,prb,preprint,superscriptaddress]{revtex4-1}
\usepackage[pdftex,colorlinks=true,linkcolor=blue,citecolor=blue,urlcolor=blue]{hyperref}
\usepackage{hyperref}
\usepackage{graphicx}
\usepackage[utf8]{inputenc}
\usepackage{placeins}

\newcommand{\figuremacro}[3]{
	\begin{figure}[tbp]
		\centering
		\includegraphics[width=#3\textwidth]{#1}
		\caption[]{#2}
		\label{fig:#1}
	\end{figure}
}

\begin{document}

\title{Soft phonon modes in rutile TiO$_2$}

\author{Bj\"orn Wehinger}
\email[]{bjorn.wehinger@unige.ch}
\affiliation{Department of Quantum Matter Physics, University of Geneva, 24, Quai Ernest Ansermet, CH-1211 Gen\`eve, Switzerland}
\affiliation{Laboratory for Neutron Scattering and Imaging, Paul Scherrer Institute, CH-5232
Villigen PSI, Switzerland}

\author{Alexe\"i Bosak}
\affiliation{European Synchrotron Radiation Facility, 71, Avenue des Martyrs, F-38000 Grenoble, France}

\author{Paweł T. Jochym}
\affiliation{Department of Computational Material Science, Institute of Nuclear Physics PAN, Cracow, Poland}

\date{\today}

\begin{abstract}
The lattice dynamics of TiO$_2$ in the rutile crystal structure was studied by a combination of thermal diffuse scattering, inelastic x-ray scattering and density functional perturbation theory. We experimentally confirm the existence of a predicted anomalous soft transverse acoustic mode with energy minimum at q = (1/2 1/2 1/4). The phonon energy landscape of this particular branch is reported and compared to the calculation. The harmonic calculation underestimates the phonon energies but despite this the shape of both the energy landscape and the scattering intensities are well reproduced. We find a significant temperature dependence in energy of this transverse acoustic mode over an extended region in reciprocal space which point to a strong role of anharmonicity in line with a substantially anharmonic mode potential-energy surface. The reported low energy branch is quite different from the ferroelectric mode that softens at the Brillouin zone center and may help explain anomalous convergence behavior in calculating TiO$_2$ surface properties and is potentially relevant for real behavior in TiO$_2$ thin films.
\end{abstract}



\maketitle

\section{Introduction}
The transition metal oxide TiO$_2$ with its mixed ionic and covalent bonding is a challenging system in condensed matter physics with many interesting applications. Its low surface energy with a particular energy landscape is, for example, very attractive for surface chemistry and photo-catalysis.
In the rutile crystal structure TiO$_2$ is an incipient ferroelectric and shows a large and strongly temperature-dependent dielectric constant. The consequent high refractive index is valuable for a diverse range of technological applications such as dielectric mirrors, antireflecting coatings and pigments. Similarly to ferroelectric perovskites, the dielectric constant increases with decreasing temperature. The incipient ferroelectric behavior and the high dielectric constant are directly linked to the lattice dynamics and they were shown to be a consequence of a low-frequency transverse optical phonon mode A$_{2u}$ \cite{lee_prb_1994}. Experimentally the phonon dispersion relation could be measured using neutron spectroscopy \cite{traylor_prb_1971}. Effects of temperature and pressure on the static dielectric constants, frequencies of Raman active modes and Gr\"uneisen parameters were addressed by capacitance measurements and Raman scattering and a particularly strong dependence on both pressure and temperature was found for the A$_{2u}$ mode and related to structural and anharmonic contributions \cite{parker_prb_1961,samara_prb_1973}.
\textit{Ab initio} lattice dynamics calculations using density functional perturbation theory (DFPT) were used to calculate phonon frequencies, Born effective charges and the dielectric permittivity \citep{lee_prb_1994, montanari_jpcm_2004, mitev_prb_2010, shojaee_jpcm_2010}. These calculations emphasize the importance of the mixed covalent ionic bonding of s and d orbitals which is linked to its large polarisability due to long range coulomb interactions between the ions. The large Born effective charges obtained could be explained by dynamical electron transfers during atomic displacements. Some of the phonon modes show a pronounced stress dependence: A significant softening of the ferroelectric A$_{2u}$ phonon mode was found \cite{montanari_jpcm_2004}. A recent theoretical work found an anomalously soft transverse acoustic (TA) mode quite separate from the A$_{2u}$ mode \cite{mitev_prb_2010}. This mode shows a remarkable stress dependence in a region of reciprocal space unexplored by experiment and softens at around q = (1/2 1/2 1/4) under isotropic expansion. The atomic displacements of all atoms for this mode at this specific momentum transfer are perpendicular to the (110) surface. They  allow for long ranged propagation of a displacement arising at the surface deep into the bulk. Under uniaxial strain the mode is predicted to soften for both expansion and compression.
The phonon frequencies are very sensitive to slight changes in the geometry and thus to the applied exchange-correlation functionals and accuracy of the pseudopotentials \cite{refson_prb_2013}.
Very recently the importance of anharmonicity in the lattice dynamics of TiO$_2$ rutile has been addressed by inelastic neutron scattering and \textit{ab initio} molecular dynamics simulations \cite{lan_prb_2015}. Temperature dependent measurements of the phonon density of states revealed a significant increase in energy of the TA phonons with temperature. The phonon potentials of this mode were reported to develop remarkable quartic terms due to a dynamic change of the hybridization between Titanium and Oxygen atoms.

Inspired by the work of Mitev \textit{et al.} \citep{mitev_prb_2010} and motivated by the fact that soft modes are very sensitive to the approximations used in theory, we experimentally address the TA mode by inelastic and diffuse x-ray scattering and compare the results to harmonic DFPT calculations. We explore an extended region of reciprocal space by thermal diffuse scattering (TDS), measure the energy of the TA phonons with mode level resolution by temperature dependent inelastic x-ray scattering (IXS) and carefully compare the phonon energy landscape and scattering intensities to the calculation. Finally we discuss details of the softening and importance of anharmonicity. 

\section{Methods}
\subsection{Experiments}
Synthetic TiO$_2$ single crystals from CrysTec GmbH (Berlin, Germany) grown by Verneuill process were used in this study. A rectangular bar with dimensions of approximately 0.2 $\times$ 0.2 $\times$ 1.5 mm$^3$ was produced from a commercially available crystal by cutting and polishing. The crystal quality was checked by high-resolution x-ray diffraction and is visible in the reconstructed planes of diffuse scattering, see Section \ref{sec:results}.

Single crystal IXS was carried out at beamline ID28 at the European Synchrotron Radiation Facility (ESRF) in Grenoble, France. The spectrometer was operated at 17.794 keV incident energy, providing an energy resolution of 3.0 meV full-width-half-maximum (FWHM). The momentum resolution was set to 0.3 (horizontal) x 0.9 (vertical) nm$^{-2}$ FWHM. Energy transfer scans were performed in transmission geometry at constant momentum transfer (Q), selected by appropriate choices of scattering angle and sample orientation. Further details of the experimental setup can be found elsewhere \cite{krisch_Springer_2007}. The temperature dependent study combines data from experiments using a closed cycle Helium cryostat (20K, 50K, 75K, 122.5K, 150K, 200K and 250K), a cryostream cooling system (105K) and no cooling device at all for the room temperature measurement.

X-ray diffuse scattering was performed on beamline ID29 at the ESRF. Monochromatic X-rays with 16 keV energy were scattered form the single crystal in transmission geometry. The sample was incrementally rotated orthogonal to the beam direction by 0.1$^{\circ}$ over a full cycle while diffuse scattering patterns were collected in shutterless mode with a single photon counting pixel detector with no readout noise and pixel size of 172 $\mu$m (PILATUS 6M detector from Dectris AG., Baden, Switzerland). The orientation matrix was defined using the CrysAlis software package (Agilent technologies, Santa Clara, USA). Reciprocal space reconstructions were prepared using locally developed software. Corrections for polarization of the incoming X-rays and for solid angle conversion associated with the planar projection were applied to the scattering intensities of individual pixels.

\subsection{Calculation}
Lattice dynamics calculations were performed in harmonic approximation using density functional perturbation theory (DFPT) \cite{gonze_prb_1997_2} as implemented in the CASTEP code \cite{clark_zkri_2005,refson_prb_2006}. 
In the calculations we employed local density approximation (LDA) with the exchange correlation functional by Perdew and Zunger \cite{perdew_prb_1981}. We used the plane-wave formalism and norm-conserving pseudopotentials of the optimised form \cite{rappe_prb_1990}. 
The LDA exchange correlation functional was chosen because the derived lattice parameters and vibrational properties are correctly described. Within the generalized-gradient approximations (PBE and PW91) the equilibrium crystal structure of TiO$_2$ is predicted to be unstable, which leads to imaginary phonon frequencies \cite{montanari_cpl_2002,refson_prb_2013}. 
We have used the pseudopotential from Mitev \textit{et al.} \cite{mitev_prb_2010} which treats five reference orbitals as valence states for Titanium and two for Oxygen. A plane wave cut-off of 1100 eV and an electronic grid sampling on a 4 $\times$ 4 $\times$ 6 Monkhorst-Pack grid ensured convergence of forces to $<$ 1 meV/\AA. The cell geometry was optimised employing the Broyden-Fletcher-Goldfarb-Shannon method \cite{pfrommer_jcp_1997} by varying lattice and internal parameters. For the cell parameters of the optimized cell we find a = b = 4.574 \AA \hspace{1pt} and c = 2.944 \AA.
Phonon frequencies and eigenvectors were computed on a 5 $\times$ 5 $\times$ 7 Monkhorst-Pack grid of the irreducible part of the Brillouin zone by perturbation calculations. Sum rules for the acoustic branches as well as the charge neutrality at the level of Born effective charges were imposed with a maximum correction of 2.0 meV at $\Gamma$. The phonon dispersion throughout the first Brillouin zone was calculated using DFPT in conjunction with Fourier interpolation with a grid spacing of 0.005 \AA$^{-1}$ in the cumulant scheme including all image force constants \cite{gonze_prb_1994,parlinski_prl_1997}. The maximum error in phonon energy from the interpolation is $<$ 1.5 meV.

Scattering intensities for inelastic and thermal diffuse scattering were calculated in first order approximation assuming the validity of both harmonic and adiabatic approximation. The formalism can be found elsewhere \cite{krisch_Springer_2007,xu_zkri_2005, wehinger_jpcm_2014}. Debye-Waller factors were calculated from a fine phonon grid with a spacing of 0.005 \AA$^{-1}$. The momentum transfer dependence of the atomic scattering factors was taken into account using an analytic function with coefficients derived from Hartree-Fock wavefunctions \cite{cromer_ac_1968}. IXS spectra are resolved in phonon energy and the intensity is sensitive to the eigenvectors. TDS intensities are not resolved in energies, but due to their strong energy dependence, they are sensitive to low energy features.

\section{Results}
\label{sec:results}
\figuremacro{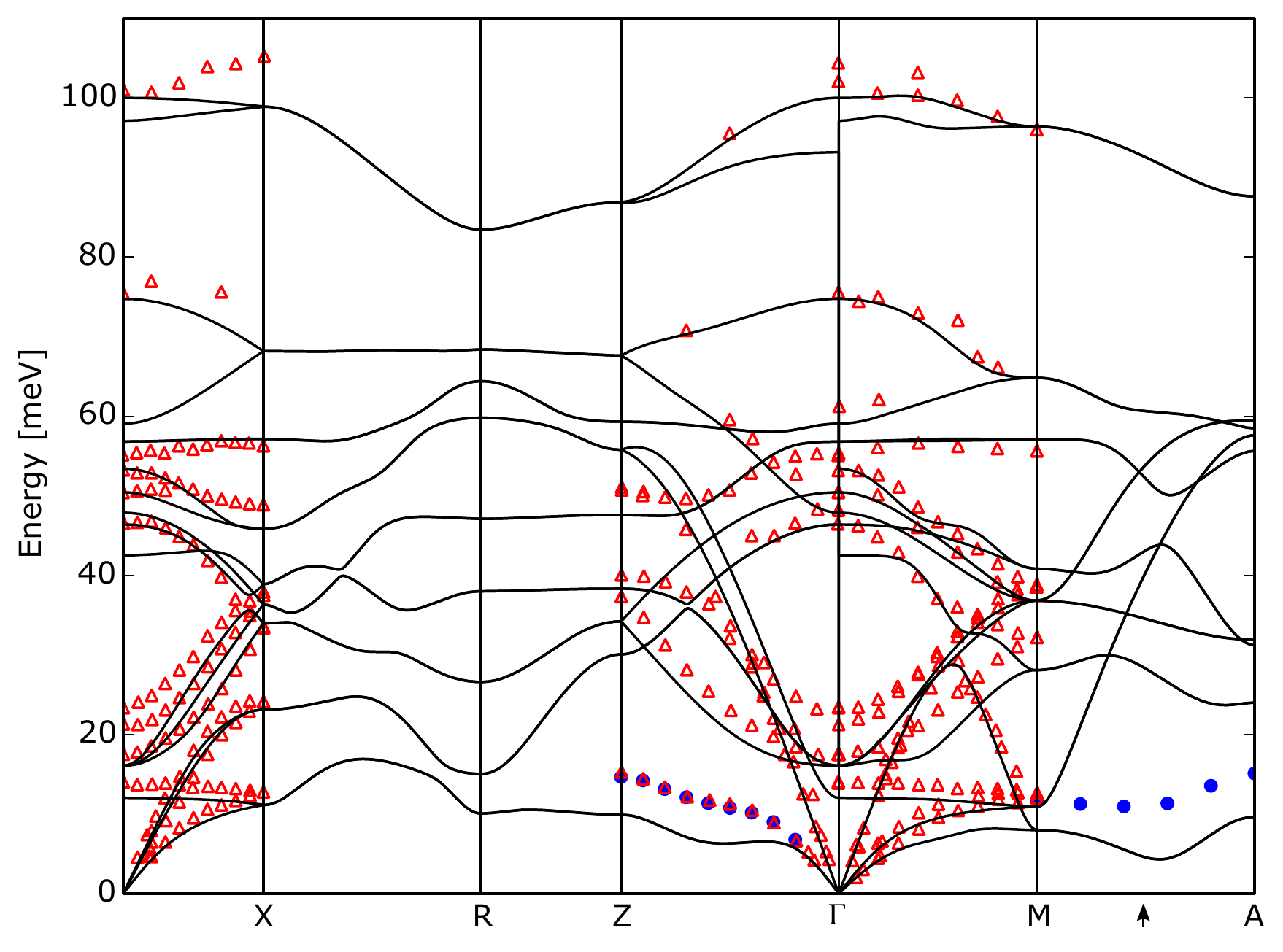}{Phonon dispersion relation of TiO$_2$ rutile, calculated by DFPT (full lines) and measured by IXS at 295 K (blue points) and INS at 300 K (red triangles) \cite{traylor_prb_1971}. The position of q = (1/2 1/2 1/4) is indicated by an arrow.}{1.0}

Phonon dispersion relations along selected high symmetry directions in the Brillouin zone are shown in Figure \ref{fig:TiO2_dispersion}. They combine the results from our calculation, IXS measurement at 295 K and previous INS data at 300 K \cite{traylor_prb_1971}. INS and IXS data are in excellent agreement, as expected. Experimental phonon energies are reproduced by the calculation within 2.5 meV or better, except for the transverse acoustic branch along $\Gamma-Z$ and $M-A$, where the discrepancy rises to nearly 5 meV. Importantly, we experimentally confirm the existence of the previously predicted low energy feature close to a reduced momentum transfer q = (1/2 1/2 1/4). IXS scans along  $M-A$ are summarized in an intensity map, see Figure \ref{fig:TiO2_IXSmap}. The dispersion clearly shows dips at Q = (2.5 2.5 0.75) and (2.5 2.5 1.25) which both correspond to q = (1/2 1/2 1/4) \footnote{Scattering intensities depend on the absolute momentum transfer. We thus report both absolute (Q) and reduced (q) values.}. The shape of the dispersion is reasonably reproduced within the harmonic (0K) DFPT calculation in spite of a significant underestimation in phonon energies.

\figuremacro{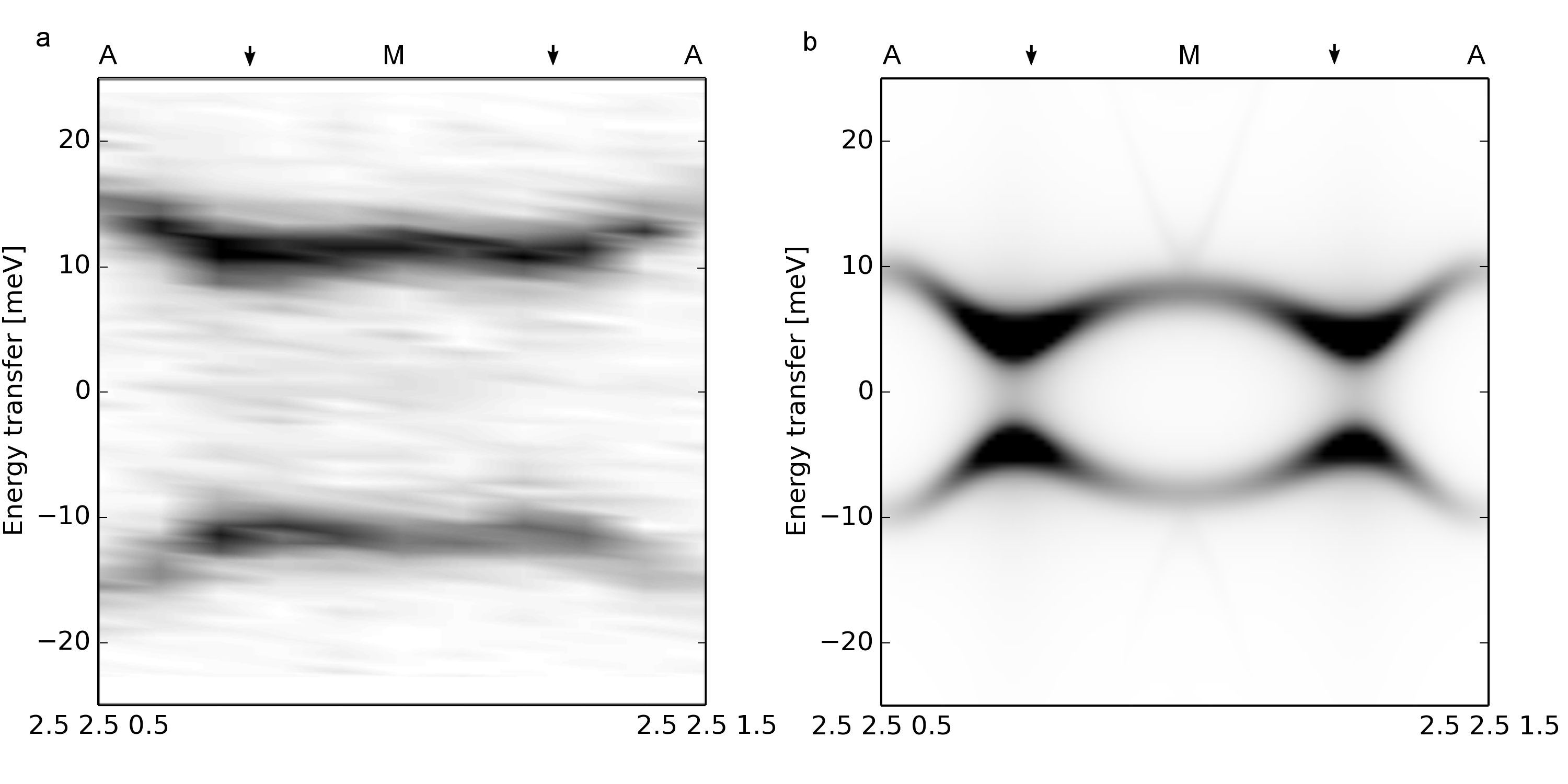}{IXS intensity of TiO$_2$ crystal as a function of energy transfer along Q = (2.5 2.5 0.5) to (2.5 2.5 1.5) ($A - M - A$), obtained from experiment at 295K (a) and calculation (b). Positions of q = (1/2 1/2 1/4) are indicated by arrows. The experimental map consists of eleven linearly spaced spectra with energy steps of 0.7 meV. The calculation was performed at 200 q-points.}{1.0}

\figuremacro{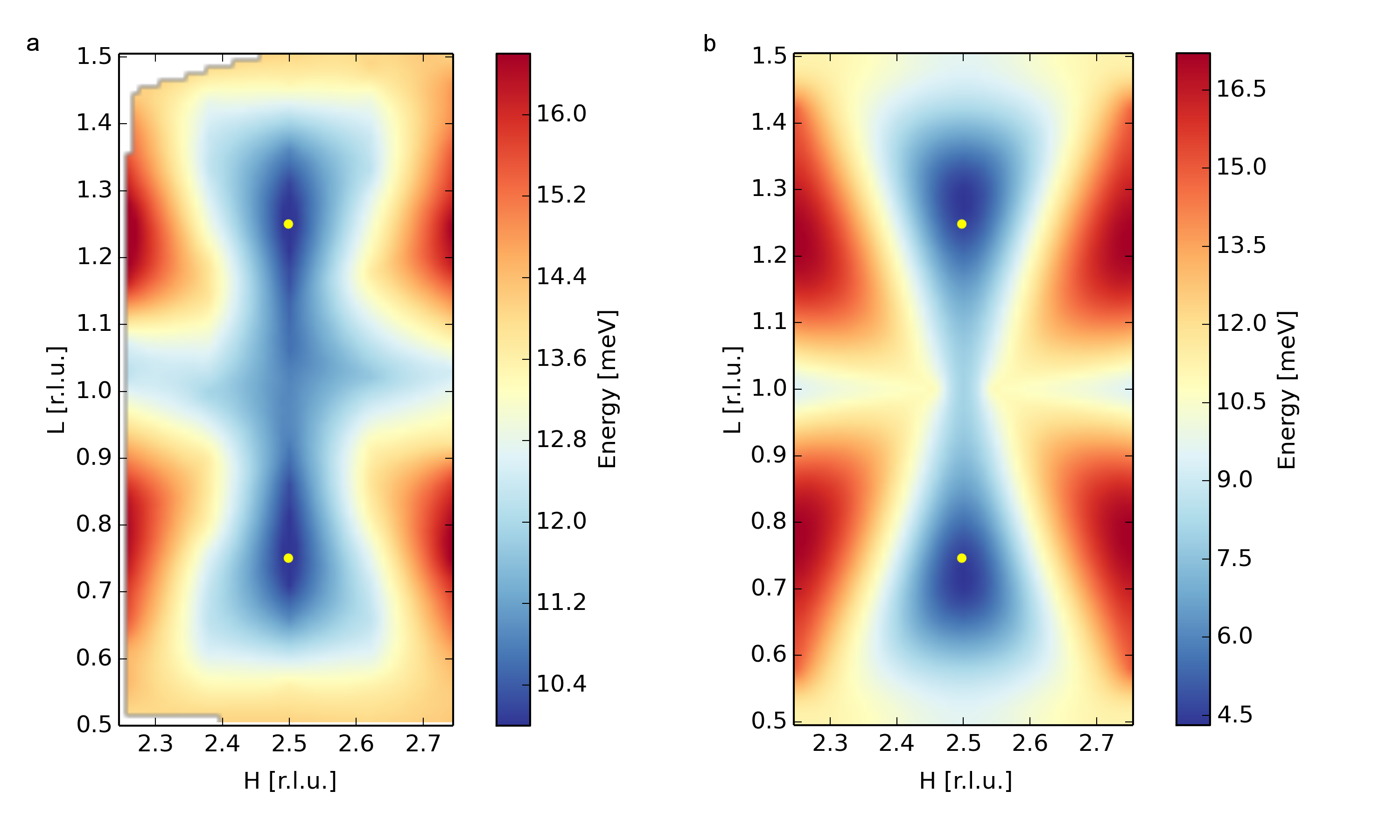}{Phonon energy landscape of the soft TA mode on the HHL plane. The phonon energies depicted in color were extracted from IXS measurements taken at 105K on 90 Q-points (a) and calculated on 5150 Q-points (b). Please note the different energy scales of the subpanels. Positions of q = (1/2 1/2 1/4) are over-plotted by dots.}{1.0}

The variation of the TA mode energy across the HHL plane is investigated by IXS and measured at a grid of Q-points on that plane. The extracted phonon energies of this particular mode are presented on a color map and compared to the calculation in Figure \ref{fig:TiO2_IXSHHL}. We note that the phonon energy landscape forms an elongated valley. The energy landscape is nicely described by the calculation along $[\xi \xi l]$. It is also well reproduced along $[h h \xi]$, except in the center of the image, where the valley appears to be a little broader (along $[0 0 \xi]$).  A closer look at the absolute energies (note the different absolute color scales in the figure) reveals that the calculated phonon energies differ by about 5 meV across the plane. Only the red areas have comparable phonon energies.

The variation in 3D reciprocal space is investigated by TDS. Scattering intensities on reconstructed high symmetry reciprocal space sections and the 3D iso-intensity distribution around Q = (2.5 2.5 1) are shown in Figure \ref{fig:TiO2_TDS} and compared to calculations. On the HK0 plane we note that the intensity distribution of TDS, which contains contributions from all phonon modes, is well reproduced by the calculation across the plane. Indeed, in this plane the calculated phonon dispersion is quite accurate. The intensity of diffuse scattering is also sensitive to the phonon eigenvectors. The excellent agreement gives high degree of confidence that the calculation properly reproduces both energies and eigenvectors across this reciprocal space section. The diffuse scattering from the TA mode along $\Gamma - M$ is particularly strong and forms streaks of high intensity for example between (0 2 0) and (2 0 0). On the HHL plane one can identify the diffuse scattering from the soft TA mode. The shape of this feature is very similar to the phonon energy landscape (Figure \ref{fig:TiO2_IXSHHL}). In fact, this mode is well separated from higher energy modes and contributes to 98 \% to the total TDS at Q = (2.5 2.5 0.75). The effect of the calculation underestimating the phonon energy is clearly noticeable: Calculated diffuse scattering intensities are significantly stronger than experimentally observed. An illustration if the iso-intensity distribution gives insight to the energy landscape in 3D, see Figure \ref{fig:TiO2_TDS} c and d. We note an elongated shape in HHL with very little variation orthogonal to it.

\figuremacro{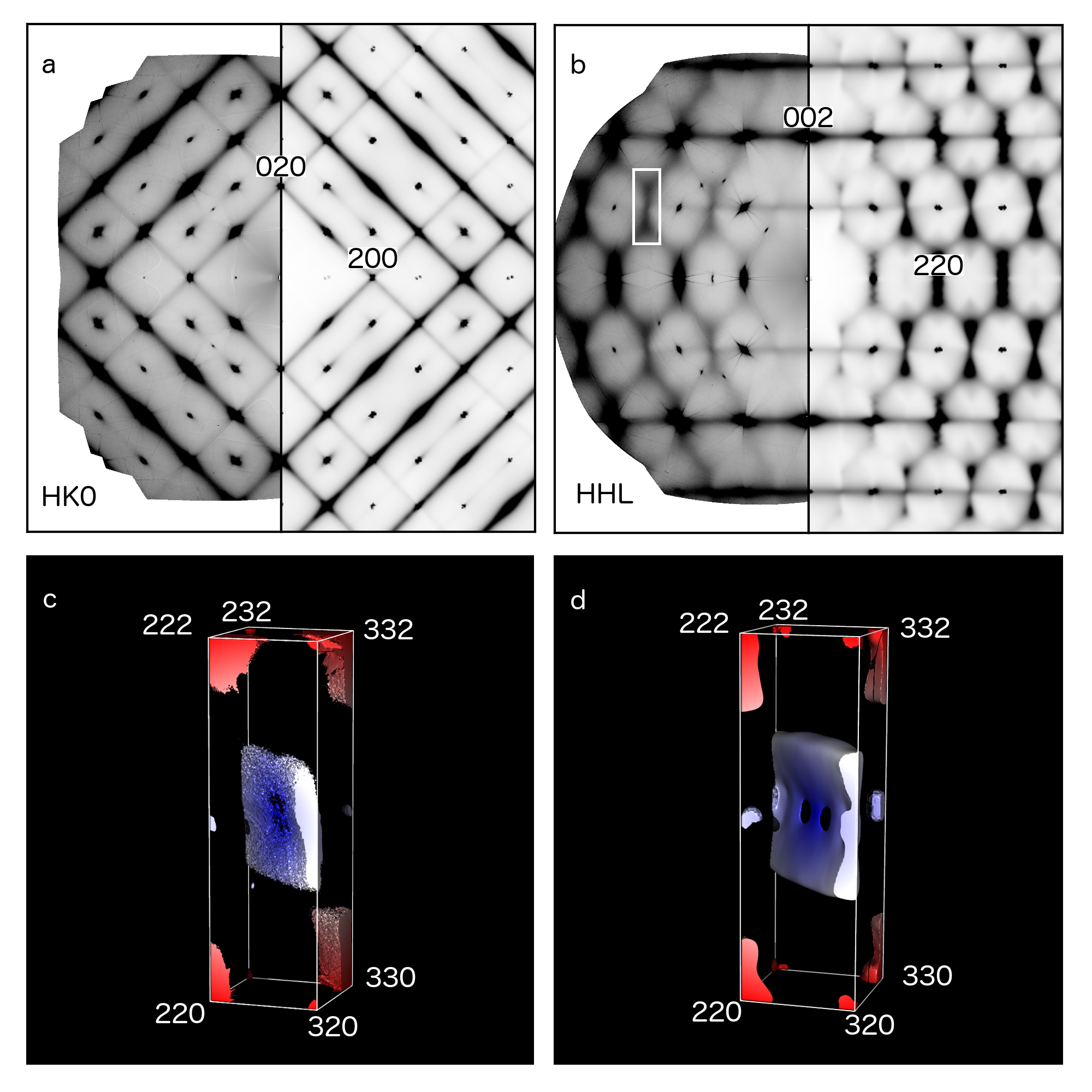}{TDS of TiO$_2$ crystal. (a) and (b) Intensity distribution across high symmetry planes in reciprocal space, reconstructed from experiment (left part of individual panels), are compared to the calculation (right part of individual panels). (c) and (d) Iso-intensity distribution of TDS in 3D around Q = (2.5 2.5 1) as obtained from experiment and calculation, respectively. The isovalue is chosen arbitrarily for best visualization. Colors indicate the distance from the (2.5 2.5 1) point.}{1.0}


The variation of selected phonon energies with temperature is shown in Figure \ref{fig:TiO2_IXStemp}. A significant temperature dependence of the soft TA mode at q = (1/2 1/2 1/4) and Z is visible. We note an almost linear increase of phonon energies with temperature. Interestingly, the dependence on Z point and q = (1/2 1/2 1/4) is very similar. For both $q$-points the calculated phonon energies are significantly too large. At q = (0 1/4 0) the temperature dependence is less pronounced and the agreement with the calculation is rather good. 

\figuremacro{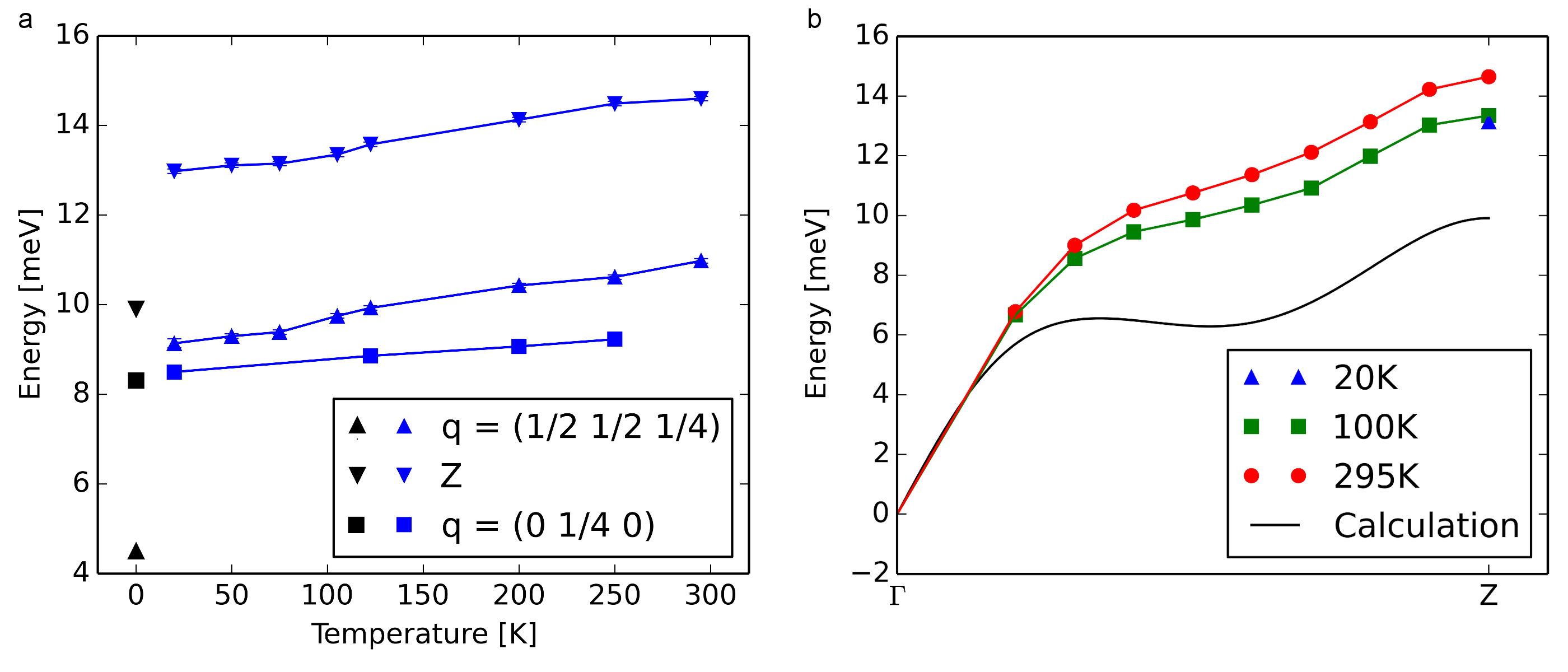}{Temperature dependence of the phonon energies of the TA mode at selected q-values. (a) Energy transfers measured by IXS (connected blue symbols) and calculated values (black symbols at zero temperature) at q = (1/2 1/2 1/4), Z and q = (0 1/4 0). (b) Dispersion along $\Gamma - Z $, measured by IXS (connected symbols, 20K data at Z point only) and calculation (solid black line).}{1.0}

The effect of anharmonicity is investigated by frozen phonon calculations at q = (1/2 1/2 1/4). The resulting phonon energy potential surfaces for different isotropic cell expansions and the corresponding displacement pattern are shown in Figure \ref{fig:TiO2_frozen} together with quadratic ($x^2$) and quartic ($x^2$ and $x^4$ terms) fits. The potential energy curves show clear evidence of anharmonicity. All curves require $x^4$ terms for a successful fit. A double-well potential forms for one percent expansion. 

\figuremacro{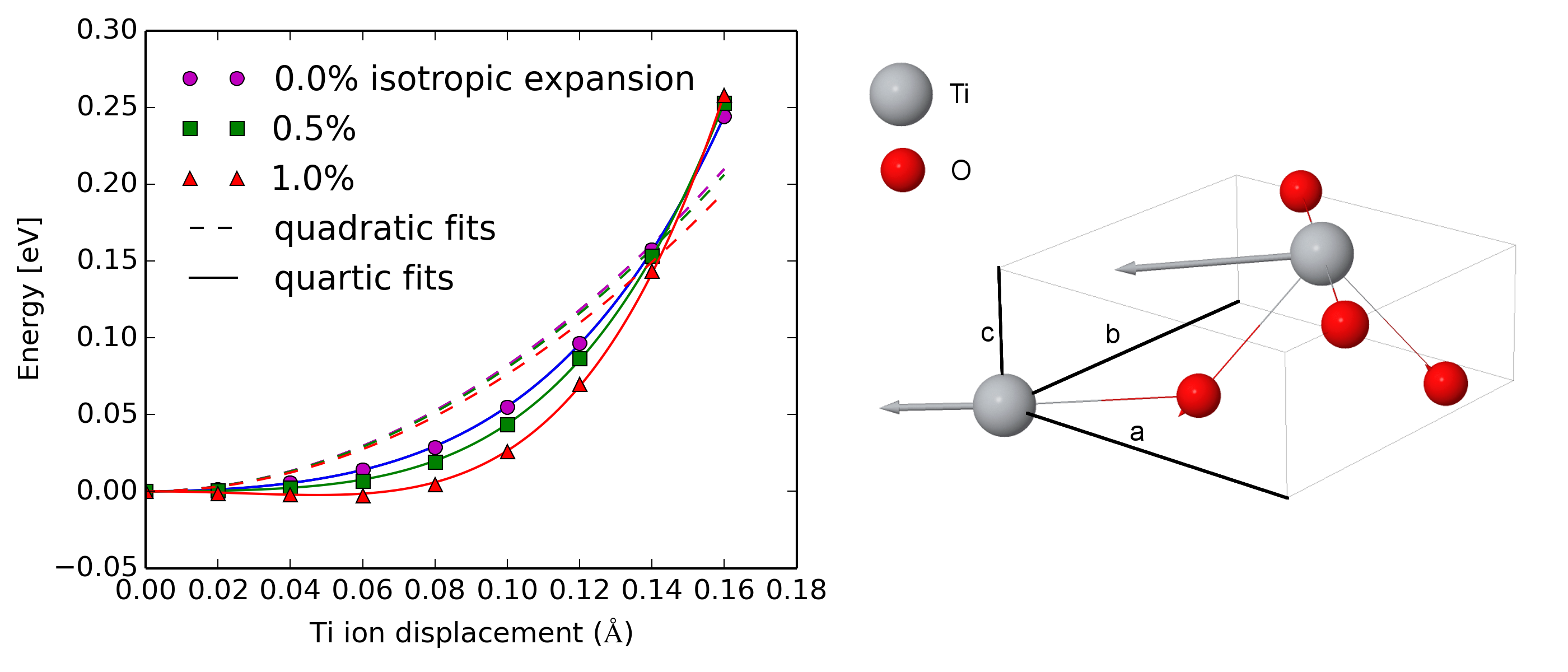}{Frozen-phonon potential-energy curves for different isotropic cell expansion (symbols \cite{mitev_prb_2010}) for the soft TA mode at q = (1/2 1/2 1/4). The mode amplitude is given for Ti ion displacement. The dashed and full lines in the corresponding color show, respectively, quadratic and quartic fits to the DFT calculation. Eigenvectors are represented within the unit cell.}{1.0}


\FloatBarrier

\section{Discussion}

The combined use of the experimental techniques TDS and IXS enables to probe extended regions in reciprocal space and thus allowed for a detailed measurement of the low energy TA phonon mode. A predicted dip structure around q = (1/2 1/2 1/4) could be confirmed. It is worth noting, that this phenomenon was entirely missed by measuring the phonon dispersion along high symmetry directions. Modern techniques, including the calculations, allow scanning across the Brillouin zone and are thus sensitive to such anomalies. Experimental results on phonon energies and scattering intensities show that the harmonic DFPT calculation reproduces dispersion and eigenvectors of most phonon branches, including its rich variety of features. The shape of phonon energy surfaces is correctly described across the Brillouin zone in spite of significantly underestimating the phonon energy of the TA mode.

The theoretical description of the lattice dynamics of TiO$_2$ in the rutile structure from first principles methods is rather challenging. The mixed covalent bonding of $s-d$ orbitals is responsible for long-ranged force constants. In order to capture subtle features in the lattice dynamics, including the incipient ferroelectric behavior and the anomaly descried here, the calculations require careful convergence tests. The present calculation profits from accurate pseudopotentials and consequently precise wave functions providing lattice parameters within 0.5 - 1.0 \% compared to the limit in LDA \cite{mitev_prb_2010}. 

Some optical phonon branches, and most importantly to this study, the lowest energy TA mode are substantially underestimated in energy. Worthy of comment is the particular sensitivity of the TA phonon mode with anomalous low energy to convergence issues and external stress. An isotropic compression of the relaxed structure by 0.5 \% brings the phonon frequencies closer to the low temperature experimental values in LDA calculations \cite{mitev_prb_2010}. 

The temperature dependent experimental data and the frozen phonon calculations show the importance of anharmonic contributions to the lattice dynamics in TiO$_2$. We find that the phonon energies at Z point and q = (1/2 1/2 1/4) increase almost linearly with temperature. Measurements of the phonon density of states as function of temperature by Lan \textit{et al.}\cite{lan_prb_2015} show that the entire TA branch stiffens with temperature. This observation is inline with our results. The change in energy at q = (0 1/4 0) is less pronounced and potentially provides important details how the phonon density of states broadens with temperature. The mode level resolution of our measurements will allow for direct comparison with temperature dependent dispersion relations. 
A proper anharmonic treatment, which is beyond the scope of this work, is required for a correct description of the TA mode. Nevertheless, even without such treatment we can extract some information from the data presented in Figure \ref{fig:TiO2_frozen}. The phonon mode potential has strong positive fourth order components. This leads to an increase of the phonon frequency for large atomic displacements (i.e. at higher temperatures) which may explain at least part of the discrepancy between purely harmonic calculation in Figure \ref{fig:TiO2_IXStemp}b and non-zero temperature measurements in the same figure. Furthermore, a rather strong dependence of fourth order components of the potential on unit cell volume indicates that thermal expansion is an important factor in the phonon frequency - temperature relationship of the investigated TA mode. 

A full quantum mechanical description of the lattice dynamics including anharmonic effects would require the computation of the multi-dimensional potential-energy surface and solving Schr\"odingers equation for the relevant degrees of freedom. \textit{Ab initio} molecular dynamics simulation followed by a mapping onto a model Hamiltonian that describes the lattice dynamics \cite{hellmann_prb_2013} could be an option. This so called 'temperature dependent effective potential technique' was successfully used to describe the temperature dependence of the phonon density of states in TiO$_2$ rutile \cite{lan_prb_2015} and phonon dispersions in VO$_2$ rutile \cite{budai_nat2014}, which is similarly affected by anharmonic contributions. The employed method should be reliable for classical nuclear motions but quantum effects need to be considered at low temperatures. Alternative methods for studying anharmonicity have been proposed, for example in References \cite{souvatzis_prl_2008, monserrat_prb_2013, errea_prb_2014}.
The presented data provide important information for benchmarking such approaches.

The harmonic DFPT calculation by Mitev \textit{et al.} \cite{mitev_prb_2010} predicted the TA mode to become soft at q = (1/2 1/2 1/4) under isotropic expansion or uniaxial strain. Our temperature dependent measurements show the opposite under thermal expansion. This is caused by the fact that even at substantial (1\%) expansion the energy difference between the global minimum and the local maximum in the potential is small in comparison with thermal energy. It suggests that the anharmonic contribution to the phonon energies at higher temperatures is much more important than the pure volume effect arising from thermal expansion. 
The open question to be addressed is what happens under application of strain. In fact, from the harmonic calculations we expect the TA mode to soften under application of external stress at small displacements. The application of uniaxial expansion could be realized by bending thin films or using tensile load. TDS and IXS studies can be performed in such conditions in grazing incidence \cite{serrano_prl_2011} or transmission geometry, respectively.

The experimental proof of the existence of the TA mode with an anomalous low energy minimum at q = (1/2 1/2 1/4) is important for the explanation of why perturbations on cleaved TiO$_2$-surfaces have an effect deep into the bulk. Our results support the proposition of Mitev \textit{et al.} \cite{mitev_prb_2010} for the anomalous slow and oscillatory convergence of thin film ('slab') models in TiO$_2$. The eigenvectors of the soft TA mode at q = (1/2 1/2 1/4) are very similar to layer displacements perpendicular to the (110) surface, see Figure \ref{fig:TiO2_frozen}. As the second layer below the surface has a structure similar to bulk TiO$_2$ \cite{cabailh_prb_2007,lindsay_prl_2005}, the high compliance of the eigenvectors allow for long-ranged propagation of surface displacements. The surface oxygen lone-pair orbitals are under-coordinated and result in strengthened Ti-O bonding and shorter Ti-O distances between the topmost pair of planes and larger distances between this plane and the next lower plane.
Depending whether a slab model consists of odd or even numbers of layers, the displacements are in or out of phase, respectively. 

From an experimental point of view this study is a benchmark for the sensitivity of diffuse x-ray scattering to particular phonon branches. In TiO$_2$ the scattering from a specific branch contributes up to 98\% to the total TDS not only close to $\Gamma$. The shape of TDS can be rapidly explored over a large volume in reciprocal space with good momentum resolution.

\section{Conclusions}
The lattice dynamics of TiO$_2$ exhibits a rich structure including an TA mode with anomalous low energy with a minimum at q = (1/2 1/2 1/4). With help of inelastic and diffuse x-ray scattering we were able to measure the energy of this mode over an extended region in reciprocal space and validate its eigenvectors by careful analysis of scattering intensities. DFPT correctly described the shape of the dispersion relations including many subtle features. Absolute phonon energies are underestimated for several branches including the TA mode of interest. It shows a significant stress dependence and is thus very sensitive to the approximations used in the calculations. 
The temperature dependent study revealed a strong temperature variation of the soft TA mode over an extended region in reciprocal space which shows the importance of anharmonic contributions to the lattice dynamics in TiO$_2$.  Our experimental observation of the soft TA mode with a minimum at q = (1/2 1/2 1/4) finally supports the explication by Mitev \textit{et al.} \cite{mitev_prb_2010} of anomalous convergence behavior in slab model calculations and real behavior in thin films.

\section*{Acknowledgment}
We would like to thank Keith Refson, Barbara Montanari and Peter Blaha for fruitful discussions and comments.

\bibliographystyle{apsrev4-1_BW}
\bibliography{TiO2_refs}

\end{document}